\documentclass{cupconf}

\newcommand{\aplt}{\,{\raise-.5ex\hbox{$\buildrel<\over\sim$}\,}}
\newcommand{\apgt}{\,{\raise-.5ex\hbox{$\buildrel>\over\sim$}\,}}

  \title[Cosmology and life]{Cosmology and life}

  \author[M.\ Livio]{M\ls A\ls R\ls I\ls O\ns   L\ls I\ls V\ls I\ls O}

  \affiliation{Space Telescope Science Institute, 3700 San Martin Drive, Baltimore, MD 21218}

\begin{document}

\maketitle

\begin{abstract}
I examine some recent findings in cosmology and their potential implications for the emergence of life in the universe. In particular, I discuss the requirements for carbon-based life, anthropic considerations with respect to the nature of dark energy, the possibility of time-varying constants of nature, and the question of the rarity of intelligent life.
\end{abstract}

\section{Introduction}

The progress in cosmology in the past few decades leads also to new insights into the global question of the emergence of intelligent life in the universe.  Here I am not referring to discoveries that are related to very localized regions, such as the detection of extrasolar planetary systems, but rather to properties of the universe at large.

In order to set the stage properly for the topics to follow, I would like to start with four observations that essentially all astronomers agree with.  These four observations \textit{define} the cosmological context of our universe.
\begin{enumerate}[(iii)]
\renewcommand{\theenumi}{\roman{enumi}}
\item \textit{Redshift}.  Ever since the observations of Vesto Slipher in 1912--1922 (Slipher 1917) and Hubble (1929), we know that the spectra of distant galaxies are redshifted.
\item Observations with the Cosmic Background Explorer have shown that, to a precision of better than 
$10^{-4}$, the cosmic microwave background is \textit{thermal}, at a temperature of 2.73~K (Mather et~al.\ 1994).
\item Light elements, such as deuterium and helium, have been synthesized in a high-temperature phase in the past (e.g.\ Gamow 1946; Alpher, Bethe \& Gamow 1948; Hoyle \& Tayler 1964; Peebles 1966; Wagoner, Fowler \& Hoyle 1967).
\item Deep observations, such as the Hubble Deep Field, have shown that galaxies in the distant universe look younger. Namely, their sizes are smaller (e.g.\ Roche et~al.\ 1996) and there is a higher fraction of irregular morphologies (e.g.\ Abraham et~al.\ 1996). This is what one would expect from a higher rate of interactions, and from ``building blocks'' of today's galaxies.
\end{enumerate}

When the above four observational facts are combined and considered together, there is no escape from the conclusion that our universe is \textit{expanding and cooling}. This conclusion is entirely \textit{consistent} with the ``hot big bang'' model. Sometimes the stronger statement, that these observations ``prove'' that there was a hot big bang, is made.  However, the scientific method does not truly produce ``proofs.''

During the past decade, deep observations with a variety of ground-based and space-based observatories have advanced our understanding of the history of the universe far beyond the mere statement that a big bang had occurred.  In particular, remarkable progress has been achieved in the understanding of the cosmic star formation history.

Using different observational tracers (e.g.\ the UV luminosity density) of star formation in high-redshift galaxies, tentative plots for the star-formation-rate (SFR) as a function of redshift have been produced (e.g.\ Lilly et~al.\ 1996; Madau et~al.\ 1996; Steidel et~al.\ 1999). There is little doubt that the SFR rises from the present to about $z\sim1$. What happens in the redshift range $z\sim1$--5 is still somewhat controversial. While some studies suggest that the SFR reaches a peak at $z\sim1$--2 and then declines slightly towards higher redshifts (e.g.\ Steidel et~al. 1999), or stays fairly flat up to $z\sim5$ (e.g.\ Calzetti \& Heckman 1999; Pei, Fall, \& Hauser 1999), others claim that the SFR continues to rise to $z\sim8$ (Lanzetta et~al.\ 2002). The latter claim is based on the suggestion that previous studies had failed to account for surface brightness dimming effects. For my present purposes, however, it is sufficient that the history of the \textit{global} SFR is on the verge of being determined (if it hasn't been determined already). A knowledge of the SFR as a function of redshift allows for the first time for meaningful constraints to be placed on the global emergence of carbon-based life.

\section{Remarks about carbon-based life}

The main contributors of carbon to the interstellar medium are intermediate-mass (1--8~M$_\odot$) stars (e.g.\ Wood 1981; Yungelson, Tutukov \& Livio 1993; Timmes, Woosley, \& Weaver 1995), through the asymptotic giant branch and planetary nebulae phases. A knowledge of the cosmic SFR history, together with a knowledge of the initial mass function (IMF; presently still uncertain for high redshift), therefore allows for a calculation of the rate of carbon production as a function of redshift (Livio 1999). For a peaked SFR, of the type obtained by Madau et~al.\ (1996), for example, the peak in the carbon production rate is somewhat delayed (by $\aplt1$~billion years) with respect to the SFR peak. The decline in the carbon production rate is also shallower for $z\aplt1$ (than the decline in the SFR), owing to the buildup of a stellar reservoir in the earlier epochs.

Assuming a ``principle of mediocrity,'' one would expect the emergence of carbon-based life to be perhaps not too far from the peak in the carbon production rate---around $z\sim1$ (for a peak in the SFR at $z\sim1$--2).  Since the timescale required to develop intelligent civilizations may be within a factor two of the lifetime of F5 to mid-K stars (the ones possessing continuously habitable zones, Kasting, Whitmore, \& Reynolds 1993; and see Section~5 below), it can be expected that intelligent civilizations have emerged when the universe was $\apgt10$~Gyr old. A younger emergence age may be obtained if the SFR does not decline at redshifts $1.2\aplt z\aplt8$ (e.g.\ Lanzetta et~al.\ 2002).

Carbon features in most anthropic arguments. In particular, it is often argued that the existence of an excited state of the carbon nucleus is a manifestation of fine-tuning of the constants of nature, that allowed for the appearance of carbon-based life.

Carbon is formed through the triple-alpha process in two steps.  In the first, two alpha particles from the unstable (lifetime $\sim10^{-16}$~s) $^8$Be. In the second, a third alpha particle is captures, via $^8$Be$(\alpha,\gamma)^{12}$C. Hoyle argued that in order for the $3\alpha$ reaction to proceed at a rate sufficient to produce the observed cosmic carbon, a resonant level must exist in $^{12}$C, a few hundred keV above the $^8$Be~+ $^4$He threshold. Such a level was indeed found experimentally (Dunbar et~al.\ 1953; Hoyle, Dunbar \& Wenzel 1953; Cook, Fowler \& Lauritsen 1957).

The question of how fine-tuned does this level need to be for the existence of carbon-based life is somewhat a matter of interpretation. The most recent work on this topic was done by Oberhammer and collaborators (e.g.\ Oberhammer, Cs\'ot\'o, \& Schlattl 2000; Cs\'ot\'o, Oberhammer, \& Schlattl 2001). These authors used a model that treats the $^{12}$C nucleus as a system of twelve interacting nucleons, and the approximate resonant reaction rate
\begin{equation}
r_{3\alpha}=3^{3/2}N^3_\alpha
\left( \frac{2\pi \hbar^2}{M_\alpha k_BT} \right)^3 
\frac{\Gamma_\gamma}{\hbar} \exp
\left(- \frac{\varepsilon}{k_BT} \right)~~.
\end{equation}
Here $M_\alpha$, $N_\alpha$ are the mass and number density of alpha particles, respectively, $\varepsilon$ is the resonance energy (in the center-of-mass frame), $\Gamma_\gamma$ is the relative width, and all other symbols have their usual meaning. Oberhammer et~al.\ introduced small variations in the strengths of the nucleon-nucleon interaction and in the fine structure constant (affecting $\varepsilon$ and $\Gamma_\gamma$), and calculated stellar models using the modified rates. They concluded that a change of more than 0.5\% in the strength of the strong interaction or more than 4\% in the strength of the electromagnetic interaction would result in essentially no production of carbon or oxygen (considering the $^{12}$C$(\alpha,\gamma)^{16}$O and $^{16}$O$(\alpha,\gamma)^{20}$Ne reactions) in any star. More specifically, a decrease in the strong-interaction strength by 0.5\% coupled with an increase in the fine structure constant by 4\% resulted in a decrease in the carbon production by a factor of a few tens in 20~M$_\odot$ stars, and by a factor of $\sim$100 in 1.3~M$_\odot$ stars.  Taken at face value, this would seem to support anthropic claims for extreme fine-tuning necessary for the emergence of carbon-based life. However, the fine-tuning appears less impressive when the following two factors are considered.
\begin{enumerate}[(iii)]
\renewcommand{\theenumi}{\roman{enumi}}
\item Calculations by Hong \& Lee (1999) show that $^{12}$C actually has an unstable resonant state of $^8$Be + $\alpha$ (as an asymmetric 3$\alpha$ linear-chain structure).
\item Calculations by Livio et~al.\ (1989) showed that shifting (artificially) the energy of the carbon resonant state by up to 0.06~MeV does not result in a significant reduction in the production of carbon. Given the results of Hong \& Lee (1999), this 0.06~MeV should be compared to the \textit{difference} between the unstable carbon state and the total energy of $^8$Be + $\alpha$, which is $\sim$0.28~MeV, and therefore not a particularly fantastic fine-tuning.
\end{enumerate}

\section{Dark energy and life}

In 1998, two teams of astronomers, working independently, presented evidence that the expansion of the universe is accelerating (Riess et~al.\ 1998; Perlmutter et~al.\ 1999). The evidence was based primarily on the unexpected faintness (by $\sim$0.25~mag) of distant ($z\sim0.5$) Type~Ia supernovae, compared to their expected brightness in a universe decelerating under its own gravity. The results favored values of $\Omega_M\sim0.3$, $\Omega_\Lambda\sim0.7$ for the matter and ``dark energy'' density parameters, respectively. Subsequent observations of the supernova SN1997ff, at the redshift of $z\simeq1.7$, strengthened the conclusion of an accelerating universe (Riess et~al.\ 2001). This supernova appeared \textit{brighter} relative to SNe in a coasting universe, as expected from the fact that at $z\sim1.7$ a universe with $\Omega_M\sim0.3$, $\Omega_\Lambda\sim0.7$ would still be in its decelerating phase. The observations of SN1997ff do not support any alternative interpretation (such as dust extinction or evolutionary effects) in which supernovae are expected to dim monotonically with redshift. Measurements of the power spectrum of the cosmic microwave background (e.g.\ Abroe et~al.\ 2002; Netterfield et~al.\ 2002; de~Bernardis et~al.\ 2002) provide strong evidence for flatness ($\Omega_M +\Omega_\Lambda=1$). When combined with estimates of $\Omega_M$ based on mass-to-light ratios, x-ray temperature of intracluster gas, and dynamics of clusters (all of which give $\Omega_M\aplt0.3$; e.g.\ Carlberg et~al.\ 1996; Strauss \& Willick 1995; Bahcall et~al.\ 2000), again a value of $\Omega_\Lambda\sim0.7$ is obtained.

Arguably the two greatest puzzles physics is facing today are:
\begin{itemize}
\item[(1)] Why is the dark energy (vacuum energy) density, $\rho_v$, so small, but not zero? (or: Why does the vacuum energy gravitate so little?)
\item[(2)] Why \textit{now}? Namely, why do we find at present that $\Omega_\Lambda\sim\Omega_M$?
\end{itemize}

The first question reflects the fact that taking graviton energies up to the Planck scale, $M_P$, would produce a dark energy density  
\begin{equation}
\rho_v\sim M_P^4\sim(10^{18}~\mathrm{GeV})^4~~,
\end{equation}
that misses the observed one, $\rho_v\sim(10^{-3}~\mathrm{eV})^4$, by more than 120 orders of magnitude. Even if the energy density in fluctuations in the gravitational field is taken only up to the supersymmetry breaking scale, $M_\mathrm{SUSY}$, we still miss the mark by a factor of 60 orders of magnitude, since $\rho_v\sim M^4_\mathrm{SUSY}\sim(1~\mathrm{TeV})^4$. Interestingly though, a scale $M_v\sim(M_\mathrm{SUSY}/M_P)M_\mathrm{SUSY}$ produces the right order of magnitude. However, while a few attempts in this direction have been made (e.g.\ Arkani-Hamed et~al. 2000), no satisfactory model has been developed. 

The second question is related to the anti-Copernican fact that $\Omega_\Lambda$ may be associated with a cosmological constant, while $\Omega_M$ declines continuously, (and in any case, $\rho_v$ may be expected to have a different time behavior from $\rho_M$), and yet the first time that we are able to measure both reliably, we find that they are of the same order.

The attempts to solve these problems fall into three general categories:
\begin{enumerate}[(iii)]
\renewcommand{\theenumi}{\roman{enumi}}
\item the behavior of ``quintessence'' fields,
\item alternative theories of gravity, and
\item anthropic considerations.
\end{enumerate}

The attempts of the first type have concentrated in particular on ``tracker'' solutions (e.g.\ Zlatev, Wang \& Steinhardt 1998; Albrecht \& Skordis 2002), in which the smallness of $\Omega_\Lambda$ is a direct consequence of the universe's old age. Generally, a uniform scalar field $\phi$, is taken to evolve according to
\begin{equation}
\ddot{\phi}+3H\dot{\phi}+V'(\phi)=0~~,
\end{equation}
where $V'(\phi)=\frac{dV}{d\phi}$ and $H$ is the Hubble parameter. The energy density of the scalar field is given by 
\begin{equation}
\rho_\phi=\frac{1}{2} \dot{\phi}^2+V(\phi)~~,
\end{equation}
and that of matter and radiation, $\rho_M$, by ($P_M$ is the pressure)
\begin{equation}
\dot{\rho}_M=-3H(\rho_M+P_M)~~.
\end{equation}
For a potential of the form
\begin{equation}
V(\phi)=\phi^{-\alpha}M^{4+\alpha}~~,
\end{equation}
where $\alpha>0$ and $M$ is an adjustable constant, $M\ll M_P$ (and a field that is initially much smaller than the Planck mass), one obtains a solution in which a transition occurs from an early $\rho_M$-dominance to a late $\rho_\phi$-dominance (with no need to fine-tune the initial conditions). Nevertheless, for the condition $\rho_\phi\sim\rho_M$ to actually be satisfied at the present time requires (Weinberg 2001) that the parameter $M$ would satisfy
\begin{equation}
M^{4+\alpha}\simeq(8\pi G)^{-1-\alpha/2}H_0^2~~,
\end{equation}
which is not easily explicable.

In order to overcome this problem, some quintessence models choose potentials in which the universe has periodically been accelerating in the past (e.g.\ Dodelson, Kaplinghat, \& Stewart 2000), so that the dark energy's dominance today appears naturally.

A very different approach regards the accelerating expansion not as being propelled by dark energy, but rather as being the result of a modified gravity. For example, models have been developed (Deffayet, Dvali, \& Gabadadze 2001), in which ordinary particles are localized on a three-dimensional surface (3-brane) embedded in infinite volume extra dimensions to which gravity can spread. The model is constructed in such a way that observers on the brane discover Newtonian gravity (four-dimensional) at distances that are shorter than a crossover scale, $r_c$, which can be of astronomical size. In one version, the Friedmann equation is replaced by
\begin{equation}
H^2+\frac{k}{a^2}=
\left( \sqrt{\frac{\rho}{3M_P^2}+\frac{1}{4r_c^2}} + \epsilon\frac{1}{2r_c^2}\right)^{\!\!2}~~,
\end{equation}
where $\rho$ is the total energy density, $a$ is the scale factor and $\epsilon=\pm1$.

In this case, the dynamics of gravity are governed by whether $\rho/M_P^2$ is larger or smaller than $1/r_c^2$. Choosing $r_c\sim H_0^{-1}$ preserves the usual cosmological results. At large cosmic distances, however, gravity spreads into extra dimensions (the force law becomes five-dimensional), and becomes weaker---directly affecting the cosmic expansion. Basically, at late times, the model has a 
self-accelerating cosmological branch with $H=1/r_c$ (to leading-order equation~(3.7) can be parameterized as $H^2-H/r_c\simeq\rho/3M_P^2$). Interestingly, it has recently been suggested that the viability of these models can be tested by Lunar Ranging experiments (Dvali, Gruzinov, \& Zaldarriaga 2002).

A third class of proposed solutions to the dark energy problems relies on anthropic selection effects, and therefore on the \textit{existence} of intelligent life in our universe.

The basic premise of this approach is that some of the constants of nature are actually random variables, whose range of values and a~priori probabilities are nevertheless determined by the laws of physics. The observed big bang, in this picture, is simply one member of an ensemble. It is further assumed that a ``principle of mediocrity'' applies, namely, we can expect to observe the most probable values (Vilenkin 1995). Using this approach, Garriga, Livio, \& Vilenkin (2000; following the original idea of Weinberg 1987) were able to show that when the cosmological constant $\Lambda$ is the only variable parameter, the order of magnitude coincidence $t_0\sim t_\Lambda\sim t_G$ (where $t_0$ is the present time; $t_\Lambda$ is the time $\Omega_\Lambda$ starts to dominate; $t_G$ is the time when giant galaxies were assembled) finds a natural explanation (see also Bludman 2000).

Qualitatively, the argument works as follows.

In a geometrically flat universe with a cosmological constant, gravitational clustering can no longer occur after redshift $(1+z_\Lambda)\sim(\rho_\Lambda/\rho_{M0})^{1/3}$ (where $\rho_{M0}$ is the present matter density). Therefore, requiring that $\rho_\Lambda$ does not dominate before redshift $z_\mathrm{max}$, at which the earliest galaxies formed, requires (e.g.\ Weinberg 1987)
\begin{equation}
\rho_\Lambda\aplt(1+z_\mathrm{max})^3\rho_{M0}~~.
\end{equation}
One can expect the a~priori (independent of observers) probability distribution $P(\rho_\Lambda)$ to vary on some characteristic scale, $\Delta\rho_\Lambda\sim\eta^4$, determined by the underlying physics. Irrespective of whether $\eta$ is determined by the Planck scale ($\sim$10$^{18}$~GeV), the grand unification scale ($\sim$10$^{16}$~GeV) or the electroweak scale ($\sim$10$^2$~GeV), $\Delta\rho_\Lambda$ exceeds the anthropically allowed range of $\rho_\Lambda$ (eq.~(3.8)) by so many orders of magnitude that it looks reasonable to assume that
\begin{equation}
P(\rho_\Lambda) = \mathit{const}~~,
\end{equation}
over the range of interest. Garriga \& Vilenkin (2001) and Weinberg (2001) have shown that this assumption is satisfied by a broad class of models, even though not automatically. With a flat distribution, a value of $\rho_\Lambda$ picked randomly (and which may characterize a ``pocket'' universe) from an interval $|\rho_\Lambda|\aplt\rho_\Lambda^\mathrm{max}$, will, with a high probability, be of the order of $\rho_\Lambda^\mathrm{max}$. The principle of mediocrity, however, means that we should observe a value of $\rho_\Lambda$ that maximizes the number of galaxies. This suggests that we should observe the largest value of $\rho_\Lambda$ that is still consistent with a substantial fraction of matter having collapsed into galaxies---in other words, $t_\Lambda\sim t_G$, as observed. In Section~2 I argued that the appearance of carbon-based life may be associated roughly with the peak in the star formation rate, $t_\mathrm{SFR}$. The ``present time,'' $t_0$, is not much different from that (in that it takes only a fraction of a stellar lifetime to develop intelligent life), hence $t_0\sim t_\mathrm{SFR}$. Finally, hierarchical structure formation models suggest that vigorous star formation is closely associated with the formation of 
galactic-size objects (e.g.\ Baugh et~al.\ 1998; Fukugita, Hogan \& Peebles 1998). Therefore, $t_G\sim t_\mathrm{SFR}$, and we obtain $t_0\sim t_G\sim t_\Lambda$.

Garriga et~al.\ (2000) further expanded their discussion to treat not just $\Lambda$, but also the density contrast at recombination, $\sigma_\mathrm{rec}$, as a random variable (see also Tegmark \& Rees 1998). The galaxy formation in this case is spread over a much wider time interval, and proper account has to be taken for the fact that the cooling of protogalactic clouds collapsing at very late times is too slow for efficient fragmentation and star formation (fragmentation occurs if the cooling timescale is shorter than the collapse timescale, $\tau_\mathrm{cool}<\tau_\mathrm{grav}$). Assuming an a~priori probability distribution of the form
\begin{equation}
P(\sigma_\mathrm{rec})\sim\sigma_\mathrm{rec}^{-\alpha}~~,
\end{equation}
Garriga et~al.\ found that ``mediocre'' observers will detect $\sigma_\mathrm{rec}\sim10^{-4}$, $t_0\sim t_G\sim t_\Lambda\sim t_\mathrm{cb}$, as observed, \textit{if} $\alpha>3$ (here the ``cooling boundary'' $t_\mathrm{cb}$ is the time after which fragmentation is suppressed).

Personally, I feel that anthropic explanations to the dark energy problems should be regarded as the \textit{last resort}, only after all attempts to find explanations based on first principles have been exhausted and failed. Nevertheless, the anthropic explanation may prove to be the correct one, if our understanding of what is truly \textit{fundamental} is lacking. A historical example can help to clarify this last statement. Johannes Kepler (1571--1630) was obsessed by the following two questions:
\begin{enumerate}[(iii)]
\renewcommand{\theenumi}{\roman{enumi}}
\item Why were there precisely six planets? (only Mercury, Venus, Earth, Mars, Jupiter and Saturn were known at his time) and
\item What was it that determined that the planetary orbits would be spaced as they are?
\end{enumerate}
The first thing to realize is that these ``why'' and ``what'' questions were a novelty in the astronomical vocabulary. Astronomers before Kepler were usually satisfied with simply recording the observed positions of the planets; Kepler was seeking a theoretical explanation. Kepler finally came up with preposterously fantastic (and absolutely wrong) answers to his two questions in \textit{Mysterium Cosmographicum}, published in 1597. He suggested that the reason for there being six planets is that there are precisely five Platonic solids. Taken as boundaries (with an outer spherical boundary corresponding to the fixed stars), the solids create six spacings. By choosing a particular order for the solids to be embedded in each other, with the Earth separating the solids that can stand upright (cube, tetrahedron, and dodecahedron) from those that ``float'' (octahedron and icosahedron), Kepler claimed to have explained the sizes of the orbits too (the spacings agreed with observations to within 10\%).

Today we recognize what was the \textit{main} problem with Kepler's model---Kepler did not understand that neither the number of planets nor their spacings are \textit{fundamental} qualities that need to have an explanation from first principles. Rather, both are the result of historical accidents in the solar protoplanetary disk. Still, it is perfectly legitimate to give an anthropic ``explanation'' for the Earth's orbital radius. If that orbit were not in the continuously habitable zone around the sun (Kasting \& Reynolds 1993), we would not be here to ask the question.

It is difficult to admit it, but our current model for the composition of the universe: $\sim$65\% dark energy, $\sim$30\% dark matter, $\sim$4\% baryonic matter, and $\sim$0.5\% neutrinos, appears no less preposterous than Kepler's model. While some version of string theories may eventially provide a 
first-principles explanation for all of these values, it is also possible, in my opinion, that these individual values are in fact not fundamental, but accidental. Maybe the only fundamental property is the fact that \textit{all the energy densities add up to produce a geometrically flat universe}, as predicted by inflation (Guth 1981; Hawking 1982; Steinhardt \& Turner 1984). Clearly, for any anthropic explanation of the value of $\Omega_\Lambda$ to be possible at all, even in principle, one requires the existence of a large ensemble of universes, with different values of $\Omega_\Lambda$. That this requirement may actually be fulfilled is precisely the concept of ``eternal inflation'' (Vilenkin 1983; Steinhardt 1983; Linde 1986; Goncharov, Linde, \& Mukhanov 1987). In most inflationary models the timescale associated with the expansion is much shorter than the decay timescale of the false vacuum phase, $\tau_\mathrm{exp}\ll\tau_\mathrm{dec}$. Consequently, the emergence of a fractal structure of ``pocket universes'' surrounded by false vacuum material is almost inevitable (Guth 2001; for a different view see e.g.\ Bucher, Goldhaber \& Turok 1995; Turok 2001).

This ensemble of pocket universes may serve as the basis on which anthropic argumentation can be constructed (even though the definition of probabilities on this infinite set is non-trivial; see e.g.\ Linde, Linde \& Mezhlumian 1995; Vilenkin 1998).

\section{Varying constants of nature?}

Another recent finding which, \textit{if confirmed}, may have implications for the emergence of life in the universe, is that of cosmological evolution of the fine structure constant $\alpha\equiv e^2/\hbar c$ (Webb et~al.\ 1999, 2001 and references therein). Needless to say, life as we know it places significant anthropic constraints on the range of values allowed for $\alpha$. For example, the requirement that the lifetime of the proton would be longer than the main sequence lifetime of stars results in an upper bound $\alpha\aplt1/80$ (Ellis \& Nanopoulos 1981; Barrow, Sandvik \& Magueijo 2001). The claimed detection of time variability was based on shifts in the rest wavelengths of redshifted UV resonance transitions observed in quasar absorption systems. Basically, the dependence of observed wave number at redshift $z$, $w_z$, on $\alpha$ can be expressed as
\begin{equation}
w_z=w_0+a_1w_1+a_2w_2~~,\end{equation}
where $a_1,a_2$ represent relativistic corrections for particular atomic masses and electron configurations, and 
\begin{eqnarray}
w_1 &= &\left(\frac{\alpha_z}{\alpha_0}\right)^2-1\\
w_2 &= &\left(\frac{\alpha_z}{\alpha_0}\right)^4-1~~.
\end{eqnarray}
Here $\alpha_0,\alpha_z$ represent the present day and redshift-$z$ values of $\alpha$, respectively. By analyzing a multitude of absorption lines from many multiplets in different ions, such as FeII and MgII transitions, in 28 absorption systems (in the redshift range $0.5\aplt z\aplt1.8$), and NiII, CrII, ZnII, and SiIV transitions in some 40 absorption systems (in the redshift range $1.8\aplt z\aplt3.5$), Webb et~al.\ (2001) concluded that $\alpha$ was \textit{smaller} in the past. Their data suggest a $4\sigma$ deviation
\begin{equation}
\frac{\Delta\alpha}{\alpha}=-0.72\pm0.18\times10^{-5}
\end{equation}
over the redshift range $0.5\aplt z\aplt3.5$ (where 
$\Delta\alpha/\alpha=\frac{\alpha_z-\alpha_o}{\alpha_o}$). It should be noted though that the data are consistent with \textit{no} variation for $z\aplt1$, in agreement with many previous studies (e.g.\ Bahcall, Sargent, \& Schmidt 1967; Wolfe, Brown, \& Roberts 1976; Cowie \& Songaila 1995).

Murphy et~al.\ (2001) conducted a comprehensive search for systematic effects that could potentially be responsible for the result (e.g.\ laboratory wavelength errors, isotopic abundance effects, heliocentric corrections during the quasar integration, line blending, and atmospheric dispersion). While they concluded that isotopic abundance evolution and atmospheric dispersion could have an effect, this was in the direction of actually amplifying the variation in $\alpha$ (to 
$\Delta\alpha/\alpha=(-1.19\pm0.17)\times10^{-5}$). The most recent results of Webb et~al.\ are not inconsistent with limits on $\alpha$ from the Oklo natural uranium fission reactor (that was active $1.8\times10^9$ years ago, corresponding to $z\sim0.1$) and with constraints from experimental tests of the equivalence principle. The former suggests $\Delta\alpha/\alpha\simeq(-0.4\pm1.4)\times10^{-8}$ (Fuji et~al.\ 2000), and the latter \textit{allows} for a variation of the magnitude observed in the context of a general dynamical theory relating variations of $\alpha$ to the electromagnetic fraction of the mass density in the universe (Livio \& Stiavelli 1998; Bekenstein 1982).

Before going any further, I would like to note that what is desperately needed right now is an independent confirmation (or refutation) of the results of Webb et~al.\ by other groups, both through additional (and preferably different) observations and via independent analysis of the data. In this respect it is important to realize that the reliability of the SNe~Ia results (concerning the accelerating universe) was enormously enhanced by the fact that two separate teams (the Supernova Cosmology Project and the High-$z$ Supernova Team) reached the same conclusion independently, using different samples and different data analysis techniques. A first small step in the direction of testing the variable $\alpha$ result came from measurements of the cosmic microwave background (CMB). A likelihood analysis of BOOMERanG and MAXIMA data, allowing for the possibility of a time-varying $\alpha$ (which, in turn, affects the recombination time) found that in general the data may prefer a smaller $\alpha$ in the past (although the conclusion is not free of degeneracies; Avelino et~al.\ 2000; Battye, Crittenden, \& Weller 2001). A second, much more important step, came through an extensive analysis using the nebular emission lines of OIII at 5007~\AA\ and 4959~\AA\ (Bahcall, Steinhardt, \& Schlegel 2003). Bahcall et~al.\ found 
$\Delta\alpha/\alpha=(-2\pm1.2)\times10^{-4}$ (corresponding to 
$|\alpha^{-1}d\alpha/dt|<10^{-13}$~yr$^{-1}$, which they consider to be a null result, given the precision of their method) for quasars in the redshift range $0.16<z<0.8$. While this result is not formally inconsistent with the variation claimed by Webb et~al., the careful analysis of Bahcall et~al.\ has cast some serious doubts on the ability of the ``Many-Multiplet'' method employed by Webb and his collaborators to actually reach the accuracy required to measure fractional variations in $\alpha$ at the $10^{-5}$ level. For example, Bahcall et~al.\ have shown that to achieve that precision, one needs to assume that the velocity profiles of different ions in different clouds are essentially the same to within 1~km~s$^{-1}$. Clearly, much more work on this topic is needed. I should also note right away that $|\Delta\alpha/\alpha|$ cannot exceed $\sim2\times10^{-2}$ at the time of nucleosynthesis, not to be in conflict with the yield of $^4$He (e.g.\ Bergstr\"om, Igury \& Rubinstein 1999). 

On the theoretical side, simple cosmological models with a varying fine structure constant have now been developed (e.g.\ Sandvik, Barrow \& Magueijo 2003; Barrow, Sandvik \& Magueijo 2002). They share some properties with Kaluza-Klein-type models in which $\alpha$ varies at the same rate as the extra dimensions of space (e.g.\ Damour \& Polyakov 1994), and with 
varying-speed-of-light theories (e.g.\ Albrecht \& Magueijo 1999; Barrow \& Magueijo 2000).

The general equations describing a geometrically flat, homogeneous, isotropic, variable-$\alpha$ universe are (Sandvik et~al.\ 2002; Beckenstein 1982; Livio \& Stiavelli 1998) the Friedmann equation (with $G=c\equiv1$)
\begin{equation}
\left(\frac{\dot{a}}{a}\right)^{\!2}=
\frac{8\pi}{3}\left[\rho_m\left(1+|\zeta_m|e^{-2\psi}\right)
+\rho_re^{-2\psi}+\rho_\psi+\rho_\Lambda\right]~~,
\end{equation}
The evolution of the scalar field varying $\alpha$ ($\alpha=\exp (2\psi)e_0^2/\hbar c$)
\begin{equation}
\ddot{\psi}+3H\dot{\psi}=-\frac{2}{w}e^{-2\psi}\zeta_m\rho_m~~,
\end{equation}
and the conservation equations for matter and radiation
\begin{eqnarray}
\dot{\rho}_m+3H\rho_m &= &0\\
\dot{\rho}_r+4H\rho_r &= &2\dot{\psi}\rho_r~~.
\end{eqnarray}
Here, $\rho_m$, $\rho_r$, $\rho_\psi$, $\rho_\Lambda$ are the densities of matter, radiation, scalar field ($=\frac{w}{2}\dot{\psi}^2$), and vacuum respectively, $a(t)$ is the scale factor ($H\equiv\dot{a}/a$), $w=\hbar c/l^2$ is the coupling constant of the dynamic Langrangian ($l$ is a length scale of the theory), and $\xi_m$ is a dimensionless parameter that represents the fraction of mass in Coulomb energy of an average nucleon compared to the free proton mass.

Equations~(4.5)--(4.8) were solved numerically by Sandvik et~al.\ (2002) and Barrow et~al.\ (2002), assuming a negative value of the parameter $\xi_m/w$, and the results are interesting both from a purely cosmological point of view and from the perspective of the emergence of life. First, the results are consistent with both the claims of a varying $\alpha$ of Webb et~al.\ (which, as I noted, badly need further confirmation) and with the more secure, by now, observations of an accelerating universe (Riess et~al.\ 1998; Perlmutter et~al.\ 1999), while complying with the geological and nucleosynthetic constraints. Second, Barrow et~al.\ find that $\alpha$ remains almost constant in the radiation-dominated era, experiences a small logarithmic time increase during the matter-dominated era, but approaches a constant value again in the $\Lambda$-dominated era. This behavior has interesting anthropic consequences. The existence of a non-zero vacuum energy contribution is now \textit{required} in this picture to dynamically stabilize the fine structure constant. In a universe with zero $\Lambda$, $\alpha$ would continue to grow in the matter-dominate era to values that would make the emergence of life impossible (Barrow et~al.\ 2001).

Clearly, the viability of all of the speculative ideas above rely at this point on the confirmation or refutation of time-varying constants of nature.

\section{Is intelligent life extremely rare?}

With the discovery of $\sim$100 massive extrasolar planets (Mayor \& Queloz 1995; Marcy \& Butler 1996, 2000), the question of the potential existence of extraterrestrial, Galactic, intelligent life has certainly become more intriguing than ever. This topic has attracted much attention and generated many speculative (by necessity) probability estimates. Nevertheless, in a quite remarkable paper, Carter (1983) concluded on the basis of the near-equality between the lifetime of the sun, $t_\odot$, and the timescale of biological evolution on Earth, $t_\ell$, that extraterrestrial intelligent civilizations are exceedingly rare in the Galaxy. Most significantly, Carter's conclusion is supposed to hold even if the conditions optimal for the emergence of life are relatively common.

Let me reproduce here, very briefly, Carter's argument. The basic, and very crucial assumption on which the argument is based is that the lifetime of a star, $t_*$, and the timescale of biological evolution on a planet around that star, $t_\ell$ (taken here, for definiteness, to be the timescale for the appearance of complex land life) are a~priori \textit{entirely independent}. In other words, the assumption is that land life appears at some \textit{random} time with respect to the main-sequence lifetime of the star. Under this assumption, one expects that generally one of the two relations $t_\ell\gg t_*$ or $t_\ell\ll t_*$ applies (the set where $t_\ell\sim t_*$ is of negligible measure for two independent quantitities). Let us examine each one of these possibilities. If \textit{generally} $t_\ell\ll t_*$, it is very difficult to understand why in the first system found to contain complex land life, the Earth-Sun system, the two timescales are nearly equal, $t_\ell\sim t_*$. If, on the other hand, \textit{generally} $t_\ell\gg t_*$, then clearly the first system we find must exhibit $t_\ell\sim t_*$ (since for $t_\ell\gg t_*$ complex land life would not have developed). Therefore, one has to conclude that \textit{typically} $t_\ell\gg t_*$, and that consequently complex land life will generally not develop---the Earth is an extremely rare exception.

Carter's argument is quite powerful and not easily refutable. Its basic assumption (the independence of $t_\ell$ and $t_*$) appears on the face of it to be solid, since $t_*$ is determined primarily by nuclear burning reactions, while $t_\ell$ is determined by biochemical reactions and the evolution of species. Nevertheless, the fact that the star is the main energy source for biological evolution (light energy exceeds the other sources by 2--3 orders of magnitude; e.g.\ Deamer 1997), already implies that the two quantities are not completely independent.

Let me first take a purely mathematical approach and examine what would it take for the condition $t_\ell\sim t_*$ to be satisfied in the Earth-Sun system \textit{without} implying that extraterrestrial intelligent life is extremely rare. Imagine that $t_\ell$ and $t_*$ are not independent, but rather that 
\begin{equation}
t_\ell/t_* = f(t_*)~~,
\end{equation}
where $f(t_*)$ is some \textit{monotonically increasing} function in the narrow range $t_*^\mathrm{min}\aplt t_*\aplt t_*^\mathrm{max}$ that allows the emergence of complex land life through the existence of continuously habitable zones (corresponding to stellar spectral types F5 to mid-K; Kasting et~al.\ 1993). Note that for a Salpeter (1955) initial mass function the distribution of stellar lifetimes behaves as 
\begin{equation}
\psi(t_*)\sim t_*~~.
\end{equation}
Consequently, if relation~(5.1) were to hold, it would in fact be the \textit{most probable} that in the first place where we encounter an intelligent civilization we would find that $t_\ell/t_*\sim1$, as in the Earth-Sun system. In other words, if we could identify some processes that are likely to produce a monotonically increasing ($t_*$; $t_\ell/t_*$) relation, then the near equality of $t_\ell$ and $t_*$ in the Earth-Sun system would find a natural explanation, with no implications whatsoever for the frequency of intelligent civilizations. A few years ago, I proposed a simple toy-model of how such a relation might arise (Livio 1999). The toy-model was based on the assumption that the appearance of land life has to await the build-up of a sufficient layer of protective ozone (Berkner \& Marshall 1965; Hart 1978), and on the fact that oxygen in a planet's atmosphere is released in the first phase from the dissociation of water (Hart 1978; Levine, Hays, \& Walker 1979). Given that the duration of this phase is inversely proportioned to the intensity of radiation in the 1000--2000~\AA\ range, a relation between $t_\ell$ and $t_*$ can be established. In fact, a simple calculation gave
\begin{equation}
t_\ell/t_*\simeq0.4(t_*/t_\odot)^{1.7}~~,
\end{equation}
precisely the type of monotonic relation needed. 

I should be the first to point out that the toy-model above is nothimg more than that---a toy model. It does point out, however, that at the very least, establishing a link between the biochemical and astrophysical timescales may not be impossible. Clearly, the emergence of complex life on Earth required many factors operating together. These include processes that appear entirely accidental, such as the stabilization of the Earth's tilt against chaotic evolution by the Moon (e.g.\ Laskar, Joutel, \& Boudin 1993). Nevertheless, we should not be so arrogant as to conclude everything from the one example we know. The discovery of many ``hot Jupiters'' (giant planets with orbital radii $\aplt0.05$~AU) has already demonstrated that the solar system may not be typical. We should keep an open mind to the possibility that biological complexity may find other paths to emerge, making various ``accidents,'' coincidences, and 
fine-tuning unnecessary. In any case, the final scientific assessment on life in the Universe will probably come from biologists and observers---not from speculating theorists like myself.

\begin{acknowledgments}
This work has been supported by Grant 938-COS191 from the Templeton Foundation.
\end{acknowledgments}
%\newpage

\end{document}